\documentclass[aps,twocolumn,showpacs,preprintnumbers,nofootinbib,prl,superscriptaddress,groupedaddress,10pt]{revtex4-1}

\makeatletter
\def\l@subsubsection#1#2{}
\def\l@subsubsubsection#1#2{}
\makeatother

\usepackage{graphicx,amssymb,amsmath,amsthm,amsfonts,epsfig,epsf}
\usepackage[usenames,dvipsnames,svgnames,table]{xcolor}
\usepackage{epstopdf}
\definecolor{darkred}{rgb}{0.5,0,0}
\usepackage{aas_macros}
\usepackage{bm}
\usepackage{dcolumn}
\usepackage[latin1]{inputenc}
\usepackage{latexsym}
\usepackage{rotating}
\usepackage{longtable}

\usepackage{enumerate}
\usepackage{tensor,multirow}
\usepackage{url}
\usepackage[linktocpage]{hyperref}

\def\be{\begin{equation}}
\def\ee{\end{equation}}
\newcommand{\beq}{\begin{eqnarray}}
\newcommand{\eeq}{\end{eqnarray}}

\def\ba{\begin{align}}
\def\ea{\end{align}}

\begin{document}
\title{Anisotropic stars as ultracompact objects in General Relativity}

\author{
Guilherme Raposo$^{1}$,
Paolo Pani$^{1}$,
Miguel Bezares$^{2}$,
Carlos Palenzuela$^{2}$,
Vitor Cardoso$^{3,4}$
}
\affiliation{$^{1}$ Dipartimento di Fisica, ``Sapienza'' Universit\`a di Roma \& Sezione INFN Roma1, P.A. Moro 5, 00185, Roma, Italy}
\affiliation{${^2}$Departament de F\'{\i}sica $\&$ IAC3, Universitat de les Illes Balears and Institut 
d'Estudis
Espacials de Catalunya, Palma de Mallorca, Baleares E-07122, Spain}
\affiliation{${^3}$ CENTRA, Departamento de F\'{\i}sica, Instituto Superior T\'ecnico -- IST, Universidade de Lisboa -- UL,
Avenida Rovisco Pais 1, 1049 Lisboa, Portugal}
\affiliation{${^4}$ Theoretical Physics Department, CERN 1 Esplanade des Particules, Geneva 23, CH-1211, Switzerland}
\begin{abstract}
Anisotropic stresses are ubiquitous in nature, but their modeling in General Relativity is poorly understood and 
frame dependent. We introduce the first study on the dynamical properties of anisotropic 
self-gravitating fluids in a covariant framework. Our description is particularly 
useful in the context of tests of the black hole paradigm, wherein ultracompact objects 
are used as black hole mimickers but otherwise lack a proper theoretical framework. 
We show that: (i)~anisotropic stars can be as compact and as massive 
as black holes, even for very small anisotropy parameters; (ii) the 
nonlinear dynamics of the $1+1$ system is in good agreement with linearized 
calculations, and shows that configurations below the maximum mass are nonlinearly stable;
(iii)~strongly anisotropic stars have vanishing tidal Love numbers in the 
black-hole limit; (iv)~their formation will usually be accompanied by gravitational-wave 
echoes at late times.
\end{abstract}

\maketitle


 \section{Introduction}
A foundational result in General Relativity~(GR) states that the maximum 
compactness of a self-gravitating, isotropic, 
spherically-symmetric object of mass $M$ and radius $R$ is $M/R=4/9$, if the 
object is composed of a perfect fluid~\cite{Buchdahl:1959zz} (we use $G=c=1$ 
units). As a corollary, under the assumptions above, the existence of exotic compact 
objects~(ECOs) of compactness arbitrarily close to that of a Schwarzschild 
black hole~(BH, with $M/R=1/2$) is ruled out. Thus, tests of the 
BH paradigm -- are the dark and massive objects that we see really BHs? -- are 
challenging to devise, impacting our ability
to {\it quantify} statements about evidence for 
BHs
\cite{Cardoso:2017cqb,Cardoso:2017njb,Barack:2018yly,Carballo-Rubio:2018jzw,Cardoso:2019rvt},
or to discover new species of compact objects.

It has been realized that Buchdahl's bound above relies strongly on the hypothesis of 
isotropy. Anisotropies in matter fields arise naturally at high 
densities~\cite{KippenhahnBook,Ruderman:1972aj,Canuto:1974gi} and may play an 
important role in the interior of compact objects. The simplest known example is 
that of a scalar field minimally coupled to gravity, which indeed gives rise 
to anisotropic pressure in boson stars~\cite{Liebling:2012fv}. Other examples 
include electromagnetic fields, fermionic fields, pion condensed phase 
configurations in neutron stars~\cite{PhysRevD.8.1260}, 
superfluidity~\cite{Carter:1998rn}, solid cores~\cite{KippenhahnBook}, etc. In 
the real world, anisotropic pressures are the rule rather than the exception.

Surprisingly, anisotropic stars in GR are poorly studied. While various 
solutions have been obtained, both in closed 
form~\cite{1974ApJ...188..657B,Letelier:1980mxb,Bayin:1982vw,Dev:2000gt,
Mak:2001eb,Dev:2003qd,Herrera:2004xc,Andreasson:2007ck} and 
numerically
\cite{1976A&A....53..283H,Doneva:2012rd,Silva:2014fca,Yagi:2015hda,Yagi:2015upa,Yagi:2016ejg,Biswas:2019gkw}, 
none arises from a consistent covariant model (see Ref.~\cite{Carloni:2017bck} for some 
progress). The lack of a proper framework 
prevents the exploration of outstanding questions associated to these objects, 
such as their stability, dynamical formation, and phenomenology. This is in 
stark contrast with the excellent knowledge on the dynamics of BHs and neutron 
stars, and is also the most important limitation in 
the study of ECOs~\cite{Cardoso:2017cqb,Cardoso:2017njb,Cardoso:2019rvt} (the only exception being boson stars which, 
however, are even less compact than the Buchdahl's bound~\cite{Liebling:2012fv} 
and do not belong to the \emph{ClePhOs} category introduced 
in Refs.~\cite{Cardoso:2017cqb,Cardoso:2017njb,Cardoso:2019rvt}).

Here we introduce a covariant and self-consistent model for 
anisotropic fluids in GR, admitting stable and well-behaved ultracompact solutions which 
we term ${\cal C}$-stars. 

For sake of simplicity we will mostly restrict our analysis to the spherically symmetric case; a covariant extension to 
the general case (without spherical symmetry) is provided in Appendix~\ref{app:beyondss}.

\section{Covariant approach to anisotropies}
Consider an anisotropic fluid with radial pressure $P_r$, tangential pressure $P_t$, and 
total energy density $\rho$, described by the stress-energy 
tensor~\cite{1974ApJ...188..657B,Doneva:2012rd}
\begin{equation}
\label{T1}
T_{\mu\nu}=\rho u_{\mu}u_{\nu}+P_r k_\mu k_\nu+P_t\Pi_{\mu\nu}\,,
\end{equation}
where $u_\mu$ is the fluid four-velocity and $k_{\mu}$ is a unit space-like 
vector orthogonal to $u_\mu$, i.e. $k^\mu k_\mu=1=-u^\mu u_\mu$, $u^\mu 
k_\mu=0$. 
Here, $\Pi_{\mu\nu}=g_{\mu\nu}+u_\mu u_\nu -k_\mu k_\nu$ is a 
projection operator onto a two-surface
orthogonal to $u^\mu$ and $k^\mu$, i.e., $u_\mu \Pi^{\mu\nu}V_\nu=k_\mu 
\Pi^{\mu\nu}V_\nu=0$ for any vector $V^\mu$.

At the center of symmetry of the fluid the anisotropy $P_r-P_t$ must 
vanish~\cite{1974ApJ...188..657B}. There is a certain degree of arbitrariness 
to satisfy this condition in a covariant fashion, the simplest possibility is
\begin{equation}
\label{eq:Pt}
P_t=P_r-{\cal C} f(\rho) k^\mu \nabla_\mu P_r\,,
\end{equation}
where $f(\rho)$ is a generic function of the density and the free constant ${\cal C}$ is 
a parameter that measures the deviation from isotropy. 
For example, for the case $f(\rho)=\rho$ considered below, ${\cal 
C}$ has the dimensions of a cubic length and Eq.~\eqref{eq:Pt} shows that the the density 
scale at which $P_t-P_r\gg P_r$ is $\rho\gg \rho_{\rm anis}$ with
\begin{equation}
 \rho_{\rm anis}\sim \frac{R}{{\cal C}}\sim 
6\times10^{15}\left(\frac{10^4}{\bar {\cal 
C}}\right)\left(\frac{R}{M}\right)\left(\frac{M}{M_\odot}\right)\frac{{\rm 
g}}{{\rm cm}^3}\,, \label{rhoanis}
\end{equation}
where $\bar{\cal C}={\cal C}/M_\odot^3$ and 
we have identified a typical lengthscale with the radius $R$.
By construction, $P_t=P_r$ at the center of static and spherically-symmetric 
objects, since $\partial_r 
P_r|_{r=0}=0$. 

By defining $\sigma:=f(\rho) k^\mu \nabla_\mu P_r$, we can write Eq.~\eqref{T1} as the 
stress-energy tensor of an 
isotropic perfect fluid plus an anisotropic contribution,
\begin{equation}
\label{T2}
T^{\mu}_{\nu}=(\rho + P_r) u^{\mu}u_{\nu}+P_r g^{\mu}_{\nu}- {\cal C} \sigma 
\Pi^{\mu}_{\nu}\,.
\end{equation}
In the spherically-symmetric case,
$u^\mu=(u^0,u^1,0,0)$, $k^\mu=(k^0,k^1,0,0)$, 
and all dynamical variables are functions of $(t,r)$ only. 
The orthogonality conditions provide two constraints on 
$k^\mu$, which is therefore completely fixed in terms of $u^\mu$.
It is straightforward to show that $\Pi^\mu_\nu= {\rm diag}(0,0,1,1)$, which 
simplifies some of the computations presented below.

\section{${\cal C}$-stars:~equilibrium configurations}
%
For static solutions, the metric can be written as 
$ds^2=-e^{\nu(r)}dt^2+(1-2m(r)/r)^{-1}dr^2+r^2\left(d\theta^2+ 
\sin^2\theta d\varphi^2\right)$.
The metric variables satisfy the standard relation, $m'(r)=4\pi r^2\rho$ and 
$\nu'(r)=2(m+4\pi r^3 P_r)/(r(r-2m))$, whereas the radial pressure satisfies a 
modified Tolman-Oppenheimer-Volkoff equation 
\begin{equation}
 P_r'(r)=-\frac{ (P_r+\rho)}{r(r-2m)}\frac{ \left(m+4\pi 
r^3 P_r\right)}{1+\frac{2}{r}{\cal C}f(\rho)\sqrt{1-\frac{2m}{r}}}\,, \label{TOV}
\end{equation}
which reduces to the isotropic case when ${\cal C}=0$.

Two equations of state for $P_r$ and $f(\rho)$ are necessary to close the system.
The simplest choice for the function $f$ would be $f(\rho)=1$, but in this model 
$P_t'$ is discontinuous at the stellar radius since $P_t'(R)\neq0$; see 
Eq.~\eqref{eq:Pt}. The simplest model that ensures 
continuity of $P_t$ and its derivative at the radius is $f(\rho)=\rho$. We focus on 
this case here, 
although other models (e.g. $f(\rho)=\rho^n$, $n>0$) give similar results. With this choice, 
Eqs.~\eqref{TOV} and~\eqref{eq:Pt} guarantee that 
$P_r=P_r'=P_t=P_t'=0$ at $r=R$.

\begin{figure}[th]
\begin{tabular}{c}
\includegraphics[width=0.49\textwidth]{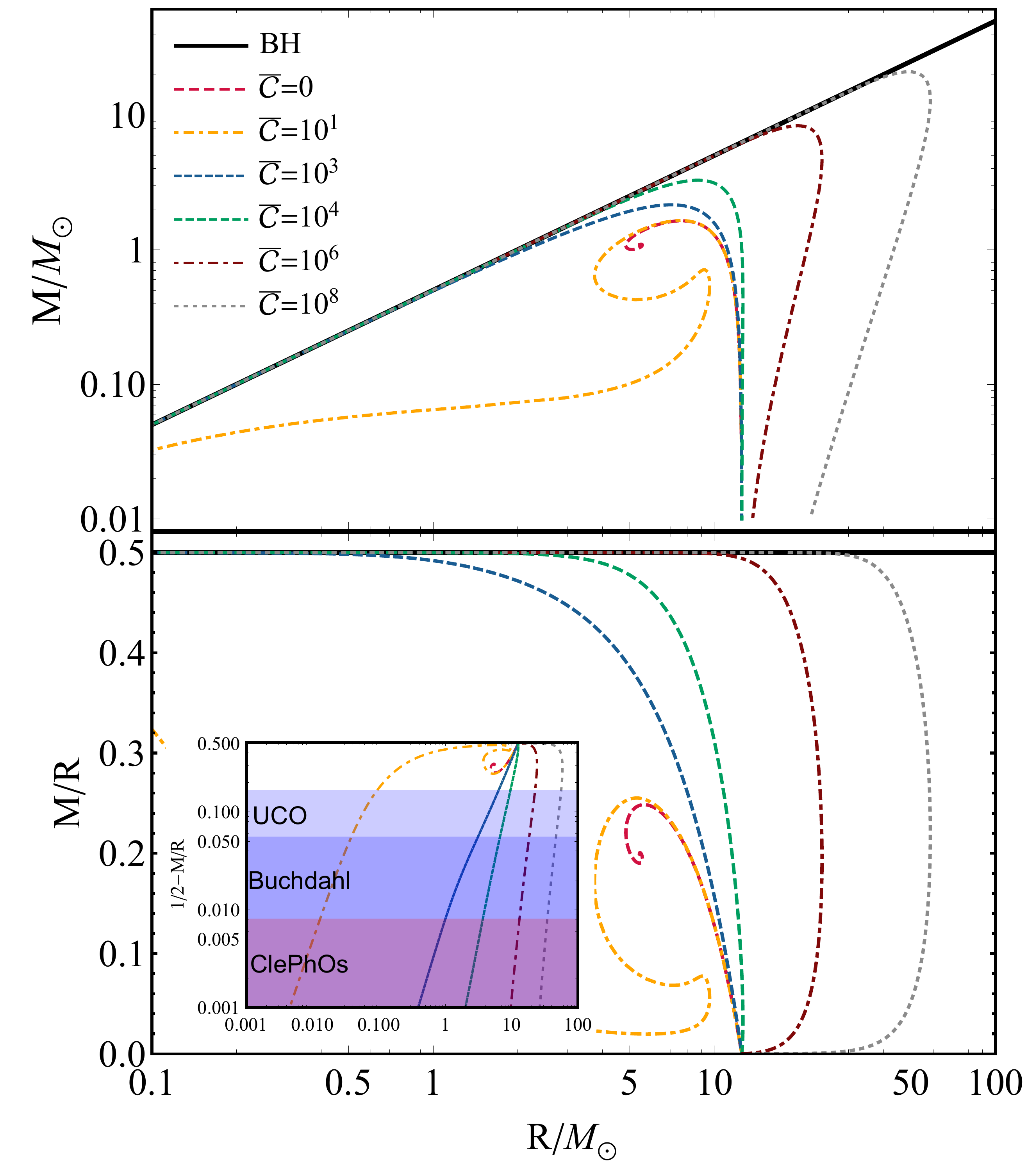}
\end{tabular}
\caption{
Mass-radius and compactness diagram for ${\cal C}$-stars with various values of the 
(dimensionless) anisotropy parameter $\bar{\cal C}$. Note 
that when $\bar{\cal C}$ is small the $M-R$ diagram shows peculiar turning 
points, which ``open up'' in the large-$\bar{\cal C}$ limit. The inset shows
the deviation $1/2-M/R$ from the compactness of a Schwarzschild BH in a logarithmic 
scale. ${\cal C}$-stars exist across the various categories (UCOs, ClePhOs)
introduced in Refs.~\cite{Cardoso:2017cqb,Cardoso:2017njb}. 
\label{fig:MR}}
\end{figure}
\begin{figure}
\includegraphics[width=0.42\textwidth]{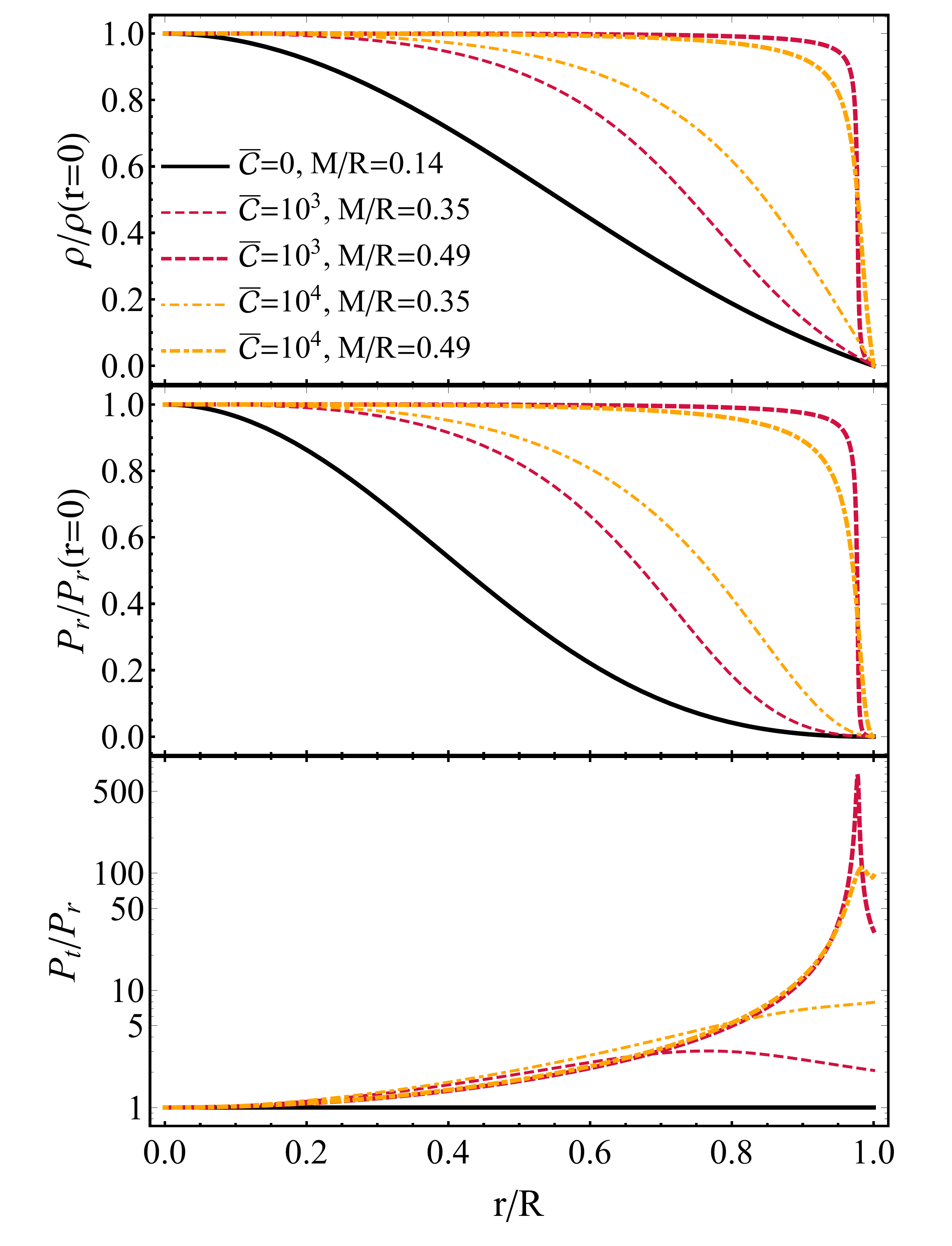} 
\caption{Energy density (top panel), radial pressure (middle panel), and 
tangential pressure (bottom panel) profiles for different 
configurations of ${\cal C}$-stars with different anisotropy parameters and 
compactness. The density and radial pressure are normalized by the corresponding values at 
the center, while the tangential pressure is normalized with the radial pressure at the 
same 
radius. The black solid line represents an isotropic configuration with $M/R=0.14$, the 
dashed thin (thick) red lines corresponds to an anisotropic configuration with $\bar{\cal 
C}=10^3$ for $M/R=0.35$ ($M/R=0.49$), and the dot-dashed thin (thick) yellow lines 
corresponds to another anisotropic configuration with $\bar{\cal C}=10^4$ for $M/R=0.35$ 
($M/R=0.49$). The results indicate that the qualitative behaviour of the fluid variables 
in this model is roughly independent of $\bar{{\cal C}}$ but depends on the compactness of 
the configuration. As $M/R\to 0.5$, the star tends to a constant-density configuration, 
while the tangential pressure profile tends to have a very sharp peak near 
the radius of the star. 
\label{anisotropy_analysis} }
\end{figure}
Remarkably, Eq.~\eqref{TOV} can be solved in closed form for a toy model of 
incompressible fluid ($\rho(r)={\rm const}$), although the solution is 
cumbersome. We focus instead on a standard 
polytropic equation of state 
$P_r=K\rho_0^\gamma$ with adiabatic index $\gamma=2$,
where $\rho_0= \rho - P_r/(\gamma-1)$ is the rest-mass density.
Our results are qualitatively the same for other standard equations of state.

The mass-radius diagram and fluid profiles of ${\cal C}$-stars are shown respectively in Fig.~\ref{fig:MR} and 
Fig.~\ref{anisotropy_analysis}, for different 
values of ${\cal C}$, related to the density scale of strong anisotropies by 
Eq.~\eqref{rhoanis}. There are several features worth highlighting: (i)~overall, ${\cal 
C}$-stars 
can be much more compact and massive than isotropic stars and their maximum 
compactness \emph{always} approaches that of a Schwarzschild BH in some region, when 
$\rho\gg\rho_{\rm anis}$; (ii)~more 
importantly, ${\cal C}$-stars exist {\it across a wide range of 
masses}, evading one of the outstanding issues with BH mimickers: in most 
theories giving rise to ECOs, these approach the BH compactness in a 
very limited range of masses, thus being unable to describe both stellar-mass 
and supermassive BH candidates across several orders of magnitude in mass.
On the other hand, ${\cal C}$-stars can do so when ${\cal C}/M^3\gg1$.
(iii)~As shown in the inset of Fig.~\ref{fig:MR}, compact configurations exceed 
the Buchdahl's limit and can even classify as
ClePhOs in the classification of~\cite{Cardoso:2017cqb,Cardoso:2017njb}.
(iv)~In general, the qualitative behaviour of the equilibrium 
solutions depends only mildly on ${\cal C}$, while it depends strongly on 
the compactness of the star. This is shown in 
Fig.~\ref{anisotropy_analysis}. Configurations with moderately low compactness display 
fluid profiles qualitatively similar to the isotropic case. However, as the compactness 
increases and approaches the black-hole compactness, the radial pressure and density 
profiles tend to constant values in the stellar interior, while the anisotropy $P_t-P_r$ 
vanishes in the core and displays a very sharp peak 
close to the radius as the star. In the $M/R\to 0.5$ limit this peak becomes 
infinitesimally thin and approaches the radius of the star, in a way reminiscent of gravastars with a thin layer of 
strongly anisotropic 
pressure~\cite{Mazur:2004fk,Visser:2003ge,Cattoen:2005he,Mazur:2015kia}.
For comparison, in Table~\ref{table2} we present data for different 
representative configurations.
\begin{table}
\resizebox{0.46\textwidth}{!}{ 
\begin{tabular}{c|c|c|c|c}
 $\bar{\cal C}$ &$\rho_c\, (\times 10^{15}{\rm gcm}^3)$ & $R/M_\odot$& $M/M_\odot$ & $\bar{\sigma}_{\rm max}$ 
\\ \hline\hline 
$10^3$ &  2.42 & 5.97 & 2.09 & 2.04 \\
$10^3$ &  3.70& 4.46 & 1.80& 4.66  \\
$10^3$ &  17.0& 1.11 &0.54& 682 \\
$10^4$ &  0.90 & 9.28 & 3.25&6.90 \\
$10^4$ &  1.20& 8.04 & 3.24&11.4 \\
$10^4$ &  3.50& 3.84 & 1.88&110
 \end{tabular}
 }
\caption{Properties of some representative ${\cal C}$-star solutions. The last column 
presents the values of $\bar{\sigma}_{\rm max}:={\rm 
max}\left\lbrace(P_t-P_r)/P_r\right\rbrace$, which gives a measurement of the maximum 
anisotropy in the interior of the star. Anisotropies are 
moderate for mildly compact configurations, whereas more compact configurations exhibit 
larger anisotropies, as also shown in the bottom panel of 
Fig.~\ref{anisotropy_analysis}.
}\label{table2}
\end{table}

(v)~When ${\cal C}\geq0$ the fluid has $P_t>0$ everywhere inside the 
star, and satisfies the weak and the strong energy 
conditions~\cite{0264-9381-5-10-011} ($\rho+P_r+2P_t\geq0$, $\rho+P_r\geq0$, 
and $\rho+P_t\geq0$), whereas very compact configurations violate the 
dominant energy condition ($\rho\geq P_r$ and $\rho\geq P_t$) near the radius, where $P_t$ attains a 
maximum and $P_t>\rho$, for very compact configurations.

\section{Radial stability}
%
For any ${\cal C}>0$ the compactness and the anisotropy grow in the high-density region,
eventually reaching the BH compactness (see inset of Fig.~\ref{fig:MR}). 
Thus, even a vanishingly small value of anisotropy parameter ${\cal C}$ can give rise to 
strongly-anisotropic quasi-Schwarzschild equilibrium 
solutions. When ${\cal C}$ is small standard 
analysis of the turning points in the mass-radius 
diagram~\cite{Shapiro:1983du} suggests that these 
configurations are unstable.
On the other hand, in the strong-anisotropy regime, the 
mass-radius relation of a ${\cal C}$-star approaches that of 
a BH already on the \emph{stable} branch. 
To test these issues, we perform a linear stability analysis of ${\cal C}$-stars under 
radial perturbations. The spacetime metric is written as 
$g_{\mu\nu}=g_{\mu\nu}^{(0)}+h_{\mu\nu}$, where $g_{\mu\nu}^{(0)}$ is the 
metric of a background ${\cal C}$-star solution and $h_{\mu\nu}={\rm 
diag}\left(H_0(r),H_2(r),0,0\right)e^{- i \omega t}$ is a small perturbation in 
Fourier space. Likewise, we expand the fluid density, pressure, and vector 
components $u^{0,1}$ and $k^{0,1}$ as $X=X_0+\delta X e^{- i \omega t}$, where 
$X_0$ collectively denotes the background quantities and $\delta X$ is the 
corresponding radial perturbation.
The orthogonality conditions on $u^\mu$ and $k^\mu$ can be used to relate $\delta 
u^0$, $\delta k^0$ and $\delta k^1$ to the remaining functions. 
Radial fluid perturbations propagate at the speed $c_s=\sqrt{\partial P_r/\partial 
\rho}$, which is always real and subluminal for these configurations. On the other hand, 
the tangential speed of sound cannot be computed in our framework since it requires 
nonspherical perturbations.

The linear system can be reduced 
to a second-order differential equation for the fluid displacement, $\xi(r)=i 
\frac{u^r}{\omega}e^{\nu/2}$. The eigenvalue problem is solved by requiring $\xi(0)=0$ 
and $\Delta P_r(R)=0$, where $\Delta P_r=\delta p+\xi \partial_r P_r$ is the 
Lagrangian variation of the pressure~\cite{Kokkotas:2000up}. This selects a discrete set of 
frequencies $\omega^2$, with $\omega^2>0$ ($\omega^2<0$) defining stable 
(unstable) modes.

\begin{figure}[bh]
\begin{tabular}{c}
\includegraphics[width=0.46\textwidth]{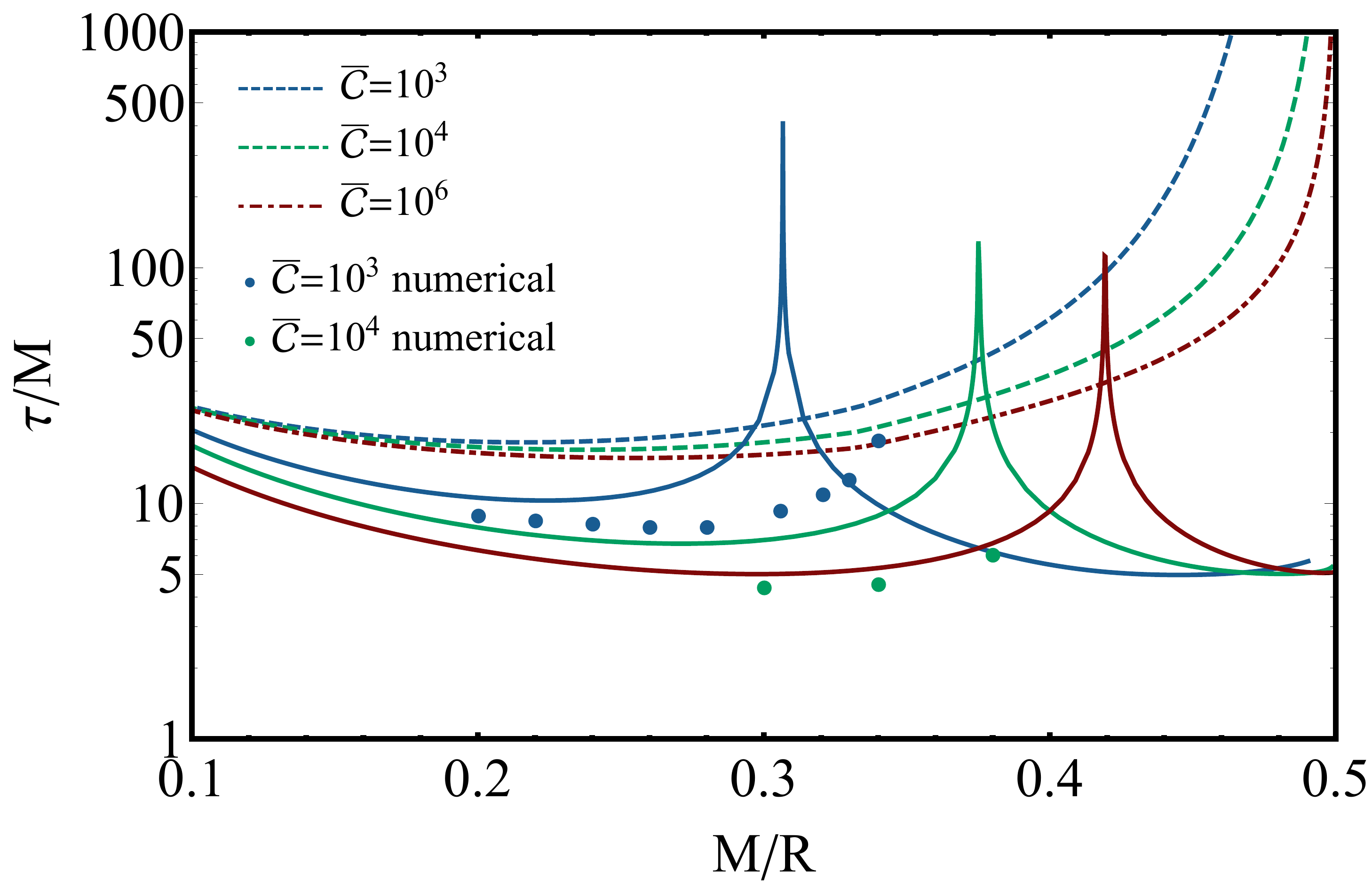}
\end{tabular}
\caption{The time scale $\tau=1/|\omega|$ for ${\cal C}$-stars as a function of the 
compactness for various values of $\bar{\cal C}$ (continuous curves). Configurations on 
the left 
of the cusps (corresponding to the zero crossing of $\omega^2$) are linearly stable, 
whereas those on the right are linearly unstable. The threshold corresponds to the 
maximum mass of the object shown in Fig.~\ref{fig:MR}. 
We also show the echo delay time~\eqref{tauecho} for these configurations (dashed 
curves). 
The markers refer to the time scale extracted from the nonlinear evolutions, which 
systematically predict more stable configurations.
\label{fig:stability}}
\end{figure}

Our results are summarized in Fig.~\ref{fig:stability}, where we show the 
fundamental modes as a function of the compactness for representative 
values of ${\cal C}$. All the expectations based on the mass-radius diagrams are 
confirmed: configurations with central density below (above) that 
corresponding to the maximum mass are linearly stable (unstable). 
Strongly-anisotropic configurations are linearly stable for 
$M/R\lesssim0.42$, while they become linearly unstable 
for higher values of the compactness.

%

\section{Evolution equations of anisotropic fluids}
We now discuss the full nonlinear theory. The conservation of stress-energy momentum
and the conservation of baryonic current
\begin{equation}\label{Tmunu_conservationb}
\nabla_\mu T^{\mu\nu} = 0 ~~~,~~~
\nabla_\mu (\rho_0 u^\mu)=0~,
\end{equation}
implies, respectively, the conservation of energy and momentum density and the conservation of mass that govern the 
fluid equations.

To write this covariant conservation law as an evolution system in spherical symmetric, one needs to split the spacetime 
tensors and equations into their space and time components by means of the $1+1$ decomposition. The line element can be 
decomposed as
\begin{equation}
ds^2 = -\alpha^2(t,r)dt^2+g_{rr}(t,r)dr^2+g_{\theta\theta}(t,r)d\Omega^2,
\end{equation}
where $\alpha$ is the lapse function, $g_{rr}$ and $g_{\theta\theta}$ are positive metric functions, and 
$d\Omega^2=d\theta^2+\sin^2\theta d\varphi^2$ the solid angle element. These quantities are defined
on each spatial foliation $\Sigma_{t}$ with normal $n_{a}=(-\alpha,0)$ and extrinsic curvature $K_{ij} \equiv 
-\frac{1}{2}\mathcal{L}_{n}\gamma_{ij}$, where $\mathcal{L}_{n}$ is the Lie derivative along $n^{a}$. 

Notice that, since $\Pi^\mu_\nu= {\rm diag}(0,0,1,1)$ in spherical symmetry, the anisotropy 
function $\sigma$ enters only in $T^{\theta}_{\theta}$ and 
$T^{\phi}_{\phi}$, being the rest of $T^\mu_\nu$ formally the same as for an isotropic fluid. The projections of this 
tensor and the baryonic current, in spherical 
symmetry, are given by
\begin{eqnarray}
D &=& \rho_0 W ~~,~ 
U = h W^2 - P_{r} ~~,~ 
S_r = h W^2 v_r, \\
S^{r}\,_{r} &=& h W^2 v^{r} v_r + P_{r} ~~,~ 
S^{\theta}\,_{\theta} = P_{r} - {\cal C}\sigma,
\end{eqnarray}
where we have defined the enthalpy $h= \rho + p = \rho_0 (1+ \epsilon) + p$ in terms of the rest mass density $\rho_0$ 
and the
internal energy $\epsilon$. Furthermore, we have defined
\begin{eqnarray}
\sigma &=& \rho (1+\epsilon)\frac{W}{\sqrt{g_{rr}}}\, \left[ 
\frac{v_{r}}{\alpha}\,\partial_{t}P_{r} + \partial_{r}P_{r} \right]\,,\\
\partial_{t}P_{r} &=& f(\alpha,g_{rr},u^{r},\partial u^
{r},\partial_{r}P_{r},\partial_{r}\rho,\sigma;
\bar{\mathcal{C}})\,,\\
u^{r}&\equiv& W v^{r} ~~~,~~~ W \equiv \frac{1}{\sqrt{1-v_r v^r}}\,.
\end{eqnarray}

It is straightforward to obtain generic evolution equations, in the sense that they do not 
depend on the specific form of the stress-energy tensor, for these projected quantities by 
projecting the conserved Eqs.~\ref{Tmunu_conservationb}. The evolved conserved 
quantities $\{D,U,S_r\}$ are not modified by the 
anisotropies. Therefore, the algorithm to convert from conserved to primitive or physical fields $\{\rho_0, \epsilon, 
P_r, v_r\}$, given an equation of state $P_r = 
P_r(\rho_0,\epsilon)$, is the same as for isotropic fluids.

Einstein's equations can be written by using 
the Z3 formulation in spherical symmetry~\cite{2010PhRvD..81d4031B}. This formulation introduces independent variables 
in order to form a first 
order evolution system. The final system must be complemented with gauge conditions for 
the lapse. We use the harmonic slicing condition $\partial_{t}\ln\alpha=-\alpha\,trK,$ 
where $trK=K_{r}^{r}+2K^{\theta}\,_{\theta}.$

Finally, the evolution system is written in balance law form~\cite{sus}
\begin{equation}
\partial_{t}{\bf U} + \partial_{i}F^{i}( {\bf U} ) = G( { \bf U})~, 
\end{equation}
where ${\bf U} 
=\{\alpha,g_{rr},g_{\theta\theta},K_{r}^{r},K^{\theta}\,_{\theta},A_{r},D_{rr}^{r},D_{r\theta}^{r},Z_{r},D,U,S_{r}\},$
is a vector containing the final set of evolution field. 
Further details on the numerical procedure and code validation are provided in Appendix~\ref{app:code}.

\begin{figure}[t]
\includegraphics[width=0.46\textwidth]{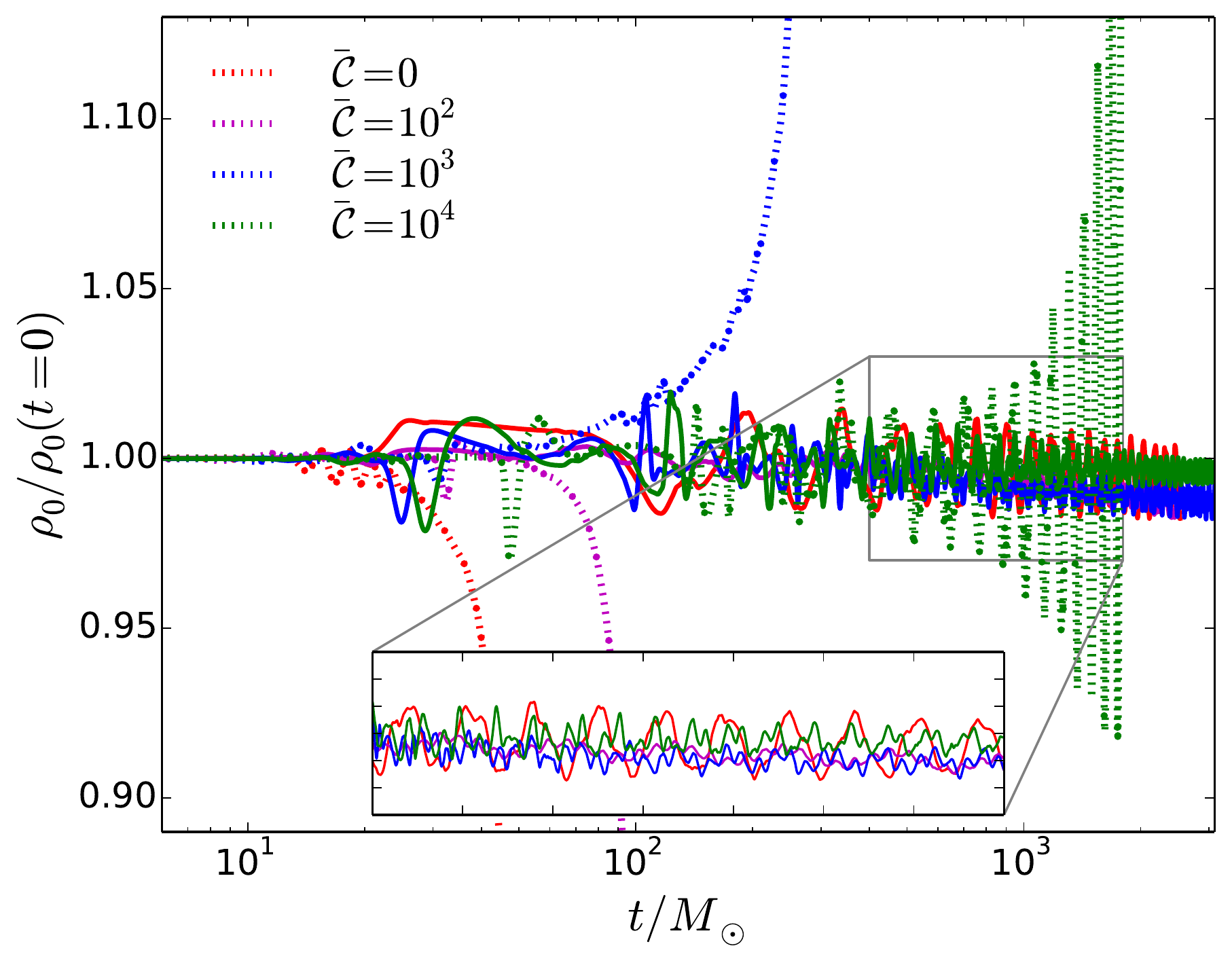} 
\caption{Central density $\rho_{0}$ as a function of time for various 
configurations in the stable (continuous curves) and in the unstable (dotted curves) 
branch.
\label{rho0} }
\end{figure}

Thus, one of our main results is that the systems of 
partial differential equations describing the anisotropic fluid and the dynamical spacetime, which are detailed above, 
is well behaved, and fully nonlinear simulations can be performed. Our 
simulations confirm the stability properties of the equilibrium configurations found 
in the previous section. Figure~\ref{rho0} displays the evolution of the central value of 
the rest-mass 
density of both stable and unstable equilibrium configurations, for different values of 
the parameter ${\cal C}$, as a function of time.
Clearly, small numerical perturbations drive unstable solutions away from 
their original configuration, whereas they remain bound for solutions in the stable 
branch.

The full nonlinear results for the timescale $\tau$ are compared with 
the linear analysis in Fig.~\ref{fig:stability}.
We find good agreement for stable configurations at moderately small compactness. 
While our code agrees very well with the linearized analysis for ${\cal C}=0$, for large compactness and large 
values of $\bar{\cal C}$, the 
nonlinear evolution shows that 
the threshold for stability is pushed to larger compactness, as compared to linear 
analysis. Furthermore, although not shown in Fig.~\ref{fig:stability}, we find indication 
that unstable configurations typically have a 
lifetime longer than the one predicted solely by linearized studies. We do not have a solid explanation for this 
discrepancy; 
it could be due to nonlinearities driving the energy to higher modes, or to other effects.
We postpone a detailed analysis for the future, but we point out that such nonlinear results are potentially exciting:
the merger of two ${\cal C}$-stars might form an ultracompact configuration which lies on 
the unstable branch, but with long lifetimes and therefore with the potential to impart 
unique signatures on the postmerger gravitational-wave~(GW) signal, some 
described below.
In addition, ${\cal C}$-stars with compactness $M/R\approx 0.4$ are nonlinearly stable. 
To the best of our knowledge, this is the first model of stable ultracompact 
objects featuring a clear photon-sphere (at $R=3M$) and which are dynamically 
well-behaved.
%
%

\section{Phenomenology of ${\cal C}$-stars}
We have just shown that ${\cal C}$-stars can be radially stable and as compact as BHs. Thus, they can mimick 
essentially all the geodesic properties of a BH~\cite{Cardoso:2019rvt}. Some smoking-gun signature will however appear 
in dynamical situations, as we now describe.

\subsection{Tidal Love numbers}
The tidal Love numbers (TLNs) define the deformability of a star immersed in an 
external field, such as the one produced by a companion in a binary~\cite{PoissonWill}. 
These quantities are particularly useful for GW 
astronomy, since they affect the late-inspiral GW signal from a coalescence and contain 
information about the nature of the merging objects~\cite{Flanagan:2007ix}. The 
prime motivation to measure TLNs is to constrain the neutron-star 
equation of state~\cite{Flanagan:2007ix,Abbott:2018exr} and to convey information on the nature of compact 
objects~\cite{Cardoso:2017cfl,Sennett:2017etc,Maselli:2017cmm,
Johnson-McDaniel:2018uvs}: in GR, the TLNs of a BH are precisely 
zero~\cite{Binnington:2009bb,Damour:2009vw,Fang:2005qq,Gurlebeck:2015xpa,
Poisson:2014gka,Pani:2015hfa}, but are nonvanishing for 
ECOs~\cite{Pani:2015tga,Uchikata:2016qku,Porto:2016zng,Cardoso:2017cfl}, being 
thus a smoking gun for ultracompact horizonless objects.

The TLNs can be computed with standard
techniques~\cite{Hinderer:2007mb,PoissonWill,Cardoso:2017cfl,Sennett:2017etc,
Maselli:2017cmm}, by studying small nonspherical (quadrupolar) deformations of a compact 
object. 
As a proof of principle, we focus on the quadrupolar \emph{scalar} 
TLNs, which are qualitatively similar to the gravitational case 
and provide the same phenomenology~\cite{Cardoso:2017cfl}.
In the large-compactness 
limit, our results are consistent with the relation
\begin{equation}
k_2^{\rm scalar} \sim a\, \bar{\cal C}^p\left(\frac{\Delta}{M}\right)^n\,,
\end{equation}
where $k_2^{\rm scalar}$ is the scalar TLN as defined in 
Ref.~\cite{Cardoso:2017cfl}, $\Delta$ is the proper distance~\cite{Maselli:2018fay} 
between $R$ 
and the Schwarzschild radius $2M$, and $a\sim{\cal O}(1)$, $p\approx 1.2$, and 
$n\approx(3-3.5)$ mildly depend on ${\cal C}$. 
Remarkably, this behavior 
is markedly different from that of other ECO models, for which the TLNs 
vanish logarithmically, $k_2\sim 1/\log(\Delta/M)$, in the BH 
limit~\cite{Cardoso:2017cfl,Maselli:2017cmm}, and shows that the TLNs of ${\cal C}$-stars 
are very small as $M/R\to1/2$. As reference, for a neutron star $k_2^{\rm scalar}\approx 
k_2^{\rm gravitational}\approx 200$ or larger~\cite{Abbott:2018exr}.

\subsection{GW echoes}
GW echoes in the post-merger GW signal from a binary coalescence are a smoking gun 
for structure at the horizon scale~\cite{Cardoso:2016rao,Cardoso:2016oxy,Ferrari:2000sr,Pani:2018flj}.
Our scope here is to simply show that perturbed ${\cal C}$-stars produce 
echoes when sufficiently compact, a more detailed analysis is left for the future. 
We 
consider a test free scalar field 
on the background of a ${\cal C}$-star. Standard spherical-harmonic and Fourier 
decomposition lead to 
$(\partial_{xx}-\partial_{tt} -V) \psi(x,t)=0$, where $x$ is the 
tortoise coordinate defined by $dr/dx=\sqrt{-g_{00}g_{11}}$, and the effective 
potential reads $V(r)=-g_{00}\left(\frac{l(l+1)}{r^2}-\frac{g_{11}'}{2 r 
g_{rr}^2}+\frac{g_{00}'}{2 r g_{11} g_{00}}\right)$, where $l=0,1,2,..$ is the harmonic 
index and $'\equiv d/dr$.
\begin{figure}[th]
\includegraphics[width=0.5\textwidth]{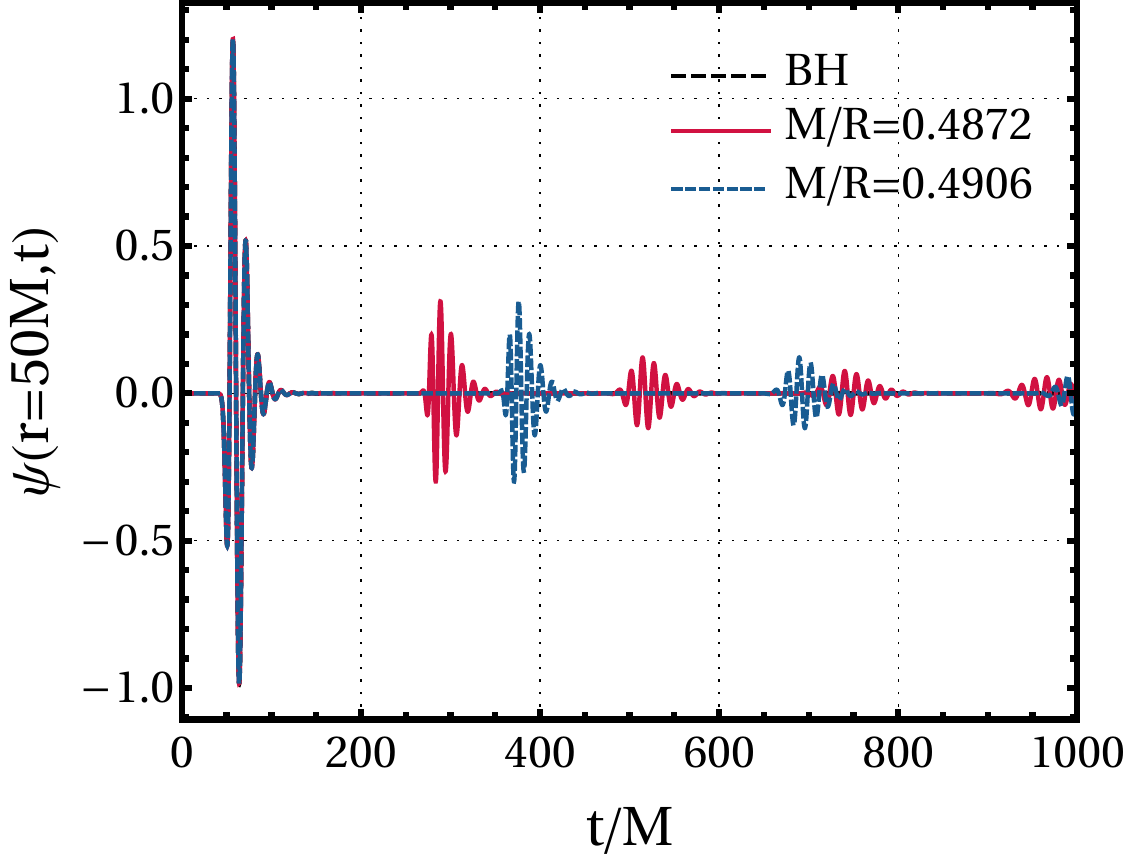} 
\caption{GW echoes from a ${\cal C}$-star with $\bar{\cal C}=10^5$ and for two 
values of the compactness. We consider quadrupolar 
scalar perturbations with an initial Gaussian profile with parameters $x_0=5M$ and 
$\sigma=4M$. The 
corresponding response of a Schwarzschild BH is shown by a continuous black 
curve for comparison. The waveforms (normalized by their peak value) are 
available online~\cite{darkGRA}. 
\label{fig:echoes}}
\end{figure}
%
Figure~\ref{fig:echoes} shows the linear response of a ${\cal C}$-star with initial 
condition $\partial_t \psi(x,0)=\exp(-(x-x_0)^2/\sigma^2)$ and $\psi(x,0)=0$.
Echoes are associated with radiation that bounces back and forth 
between the object and the photon-sphere~\cite{Cardoso:2014sna}, slowly 
leaking to infinity through wave tunneling~\cite{Cardoso:2016rao,Cardoso:2016oxy}. Thus,
the time delay between echoes roughly corresponds to twice the light crossing 
time from the center of the star to the photon 
sphere~\cite{Cardoso:2016rao,Cardoso:2016oxy,Pani:2018flj},
\begin{equation}
 \tau_{\rm 
echo}=\int_{0}^{3M} \frac{dr}{\sqrt{e^{\nu}(1-2m/r)}}\,.\label{tauecho}
\end{equation}
Interestingly, this delay time is typically 
dominated by the the travel time \emph{within} the star, not by the the Shapiro 
delay factor $\sim 
\log(1-2M/R)$ near the surface~\cite{Cardoso:2016oxy}. This 
property is akin to (isotropic) ultracompact stars near the Buchdahl's 
limit~\cite{Cardoso:2017njb,Pani:2018flj} and to certain phenomenological 
models~\cite{Barausse:2018vdb} considered in the past.

\section{ Discussion}
%
To summarize, we introduced a covariant framework to study anisotropic stars in GR, whose resulting evolution system is 
well-posed. Thus, relativistic anisotropic fluids can be explored in full-blown nonlinear 
evolutions, as we hope to do in the near future.

Our results are exciting: ${\cal C}$-stars provide a prototypical model for 
ultracompact objects, of immense utility in the quest to quantify the evidence
for BHs and in identifying possible smoking guns for new 
physics~\cite{Cardoso:2017njb,Cardoso:2017cqb,Carballo-Rubio:2018jzw}.
Such configurations can be metastable and display the whole 
phenomenology recently predicted for ultracompact horizonless objects. In 
particular, they can be as massive and compact as BHs, have vanishingly small 
TLNs~\cite{Cardoso:2017cfl}, and produce 
GW echoes~\cite{Cardoso:2016rao,Cardoso:2016oxy} when perturbed (evading recent 
results~\cite{Urbano:2018nrs}, due to anisotropy).

It is intriguing to notice that ${\cal C}$-stars share many key properties 
with gravastars~\cite{Mazur:2004fk,Cattoen:2005he}, although the dynamics of the 
latter lacks a solid theoretical framework (see~\cite{Beltracchi:2018ait} for recent 
progress). Thus, ${\cal C}$-stars 
might serve as an effective model for semiclassical corrections near the 
horizon, as predicted in other contexts~\cite{Mazur:2015kia,Carballo-Rubio:2017tlh}. 
Some of these models are nonperturbative in the Planck length $\ell_P$, and they would 
predict
${\cal C}/M^3\sim M/\ell_P\sim 10^{38}$ for $M=M_\odot$, which motivates the 
strong-anisotropy regime explored here. It is also likely that the magnitude of 
anisotropies grows with the 
compactness of the object. Anisotropic effects might become stronger during the 
merger and an ordinary neutron star might ``anisotropize'' dynamically.

We have worked with a very crude toy model. Generalizations include, for example, models 
with $f(\rho)=\rho^n$; preliminary results for $n>1$ show that stars in the stable 
branch are even more compact than the models presented here.
We are tempted to conclude that there are very generic models of anisotropy which
lead to the same phenomenology as the one we report.

Finally, we focused on the nonlinear dynamics in the spherically symmetric 
case; extensions of our covariant formalism to less symmetric configurations 
and on simulations of binary ${\cal C}$-stars are ongoing, based on the general covariant
framework presented in the Appendix~\ref{app:beyondss}. 
This is particularly interesting in light of our results: the mass-radius 
diagram of ${\cal C}$-stars suggests that, for (say) $\bar{\cal C}=10^{8}$, two 
merging ${\cal C}$-stars with equal mass $M\approx11\,M_\odot$ 
(compactness $M/R\approx 0.18$) might give rise to a \emph{stable} ${\cal 
C}$-star near the maximum mass, $M_f\approx 21\, M_\odot$ and with 
compactness $M_f/R \approx 0.43$, being thus a viable candidate for an ${\rm ECO}+{\rm 
ECO}\to{\rm ECO}$ coalescence.

\noindent{\bf{\em Acknowledgments.}}
%
We thank Sante Carloni, Kento Yagi, and Stoytcho Yazadjiev for useful comments and feedback.
We are grateful to Kinki University in Osaka for kind hospitality while this work begun.
The authors thank the Yukawa Institute for Theoretical Physics at Kyoto University and 
the Galileo Galilei Institute in Florence, where this work was completed during the 
YITP-T-18-05 on ``Dynamics in Strong Gravity Universe'' and during the ``Fundamental 
Physics with LISA'' workshop, respectively.
PP acknowledges financial support provided under the European Union's H2020 ERC, Starting Grant agreement
no.~DarkGRA--757480 and the kind hospitality of the
Universitat de les Illes Balears, where part of this work has been performed.
CP and MB acknowledge support from the Spanish Ministry of Economy and Competitiveness grants AYA2016-80289-P 
(AEI/FEDER, UE) and AYA2017-82089-ERC. CP also acknowledges support from the Spanish Ministry of Education and Science 
through a Ramon y Cajal grant. MB would like to thank CONICYT Becas Chile (Concurso Becas de Doctorado en el 
Extranjero) for financial support.
MB also acknowledge the support of the PHAROS COST Action (CA16214).
V.C. acknowledges financial support provided under the European Union's H2020 ERC 
Consolidator Grant ``Matter and strong-field gravity: New frontiers in Einstein's 
theory'' grant agreement no. MaGRaTh--646597.
This project has received funding from the European Union's Horizon 2020 research and innovation programme under the 
Marie Sklodowska-Curie grant agreement No 690904.
We acknowledge financial support provided by FCT/Portugal through grant PTDC/MAT-APL/30043/2017.
We acknowledge the SDSC Comet and TACC Stampede2 clusters through NSF-XSEDE Award Nos. PHY-090003,
as well as MareNostrum and the technical support provided by Barcelona Supercomputing Center (AECT-2018-1-0003).
The authors would like to acknowledge networking support by the GWverse COST Action 
CA16104, ``Black holes, gravitational waves and fundamental physics.''
We acknowledge support from the Amaldi Research Center funded by the 
MIUR program ``Dipartimento di Eccellenza''~(CUP: B81I18001170001).
%

\appendix

\section{General covariant framework beyond spherical symmetry}\label{app:beyondss}
In this appendix we generalize the covariant framework for anisotropic fluids in GR without any special symmetry. In the 
most general case, the pressure can be 
different along three generic spatial directions, identified by a triad of unit 
space-like vectors, $k_{\mu}$, $\xi_{\mu}$ and $\eta_{\mu}$. The anisotropies can be 
defined in terms of deviations $\sigma_1$, $\sigma_2$ and $\sigma_3$ from a reference 
isotropic pressure function, $P_{\rm ISO}$.
The stress-energy tensor describing the resulting 
anisotropic fluid can be written as
\begin{align}
\label{T1b}
T_{\mu\nu}&=T^{\rm ISO}_{\mu\nu}+\sigma_1k_\mu k_\nu+\sigma_2\xi_\mu \xi_\nu+\sigma_3 
\eta_\mu \eta_\nu\,,
\end{align}
where we have defined the stress-energy tensor of an ordinary isotropic fluid as
\begin{align}
\label{TISO}
T^{\rm ISO}_{\mu\nu}&=(\rho +P_{\rm ISO}) u_{\mu}u_{\nu}+g_{\mu\nu}P_{\rm ISO}\,,
\end{align}
with $u_\mu$ being the usual fluid four-velocity. 

The unit vectors $k_{\mu}$, $\xi_{\mu}$ and $\eta_{\mu}$ are orthogonal to each other and 
to 
$u_\mu$, i.e. they are constrained by the nine
conditions: $k^\mu k_\mu=\xi^\mu \xi_\mu=\eta^\mu \eta_\mu=1$, and $u^\mu 
k_\mu=u^\mu \xi_\mu=k^\mu \xi_\mu=k^\mu \eta_\mu=\xi^\mu \eta_\mu=u^\mu \eta_\mu=0$. 
These conditions can be used to fix nine out of the $3\times4$ components of $k_{\mu}$, 
$\xi_{\mu}$ and $\eta_{\mu}$. The remaining three components can be arbitrarily fixed 
without loss of generality, since they are associated with the translation of the origin 
of the frame identified by the triad of pressure vectors.

Using the above conditions, we can define the projections
\begin{align}
&u^\mu T_{\mu\nu}=\rho u_\nu \,,\\
&k^\mu T_{\mu\nu}=\left(P_{\rm ISO}+\sigma_1\right) k_\nu= P_1 k_\nu\,,\\
&\xi^\mu T_{\mu\nu}=\left(P_{\rm ISO}+\sigma_2 \right)\xi_\nu= P_2 \xi_\nu\,,\\
&\eta^\mu T_{\mu\nu}=\left(P_{\rm ISO}+\sigma_3 \right)\eta_\nu= P_3 \eta_\nu\,,
\end{align}
where we defined $P_i:=P_{\rm ISO}+\sigma_i$. Thus, in this notation each vector is 
related with a specific direction of anisotropy. When $\sigma_1=\sigma_2=\sigma_3=0$ the 
stress-energy tensor above reduces to the usual stress-energy tensor for an isotropic 
perfect fluid.

In analogy to the spherically symmetric case, we provide three equations of state for 
$P_1$, $P_2$ and $P_3$ of the form
\begin{align}
\label{eq:tangencialp1}
\sigma_1&=P_1-P_{\rm ISO}={\cal C}_1 f_1(\rho) k^\mu \nabla_\mu \rho\,,\\
\sigma_2&=P_2-P_{\rm ISO}={\cal C}_2 f_2(\rho) \xi^\mu \nabla_\mu \rho\,,\\
\sigma_3&=P_3-P_{\rm ISO}={\cal C}_3 f_3(\rho) \eta^\mu \nabla_\mu 
\rho\,,\label{eq:tangencialp3}
\end{align}
where the free constants ${\cal C}_1$, ${\cal C}_2$ and ${\cal C}_3$ are generically 
dimensionful. At variance with the main text, we have defined the equations of state in 
terms of covariant derivatives of the density, since the latter is unique.

Let us now show that the general framework reduces to the spherically symmetric case 
considered in the main text. In spherical symmetry, the angular components of the vectors 
must vanish. In this case, $k_\mu=(k_0,k_1,0,0)$ and the other two vectors are 
identically zero, $\xi_\mu=0$ and $\eta_\mu=0$. 
Equations~\eqref{eq:tangencialp1}-\eqref{eq:tangencialp3} then take the form
\begin{align}
\label{eq:tangencialp2}
P_1&=P_{\rm ISO}+{\cal C}_1 f_1(\rho) k^\mu \nabla_\mu \rho\,,\\
P_2&=P_{\rm ISO}\,,\\
P_3&=P_{\rm ISO}\,,
\end{align}
and therefore $\sigma_2=\sigma_3=0$. 
The spherically-symmetric case is recovered by defining $P_t=P_{\rm ISO}$, 
$P_t=P_r-{\cal C}_1 f_1(\rho) k^\mu \nabla_\mu \rho$, and $f_1(\rho)=f(\rho) \partial 
P_r/\partial\rho$, to account for the different definition in the equation of state.
%

\section{Code validation}\label{app:code}

The numerical code used in this work is a simple extension of the one presented in 
Ref.~\cite{sus} to study fermion-boson stars. The spatial discretization of spacetime 
fields is performed using a third-order accurate Finite Volume method 
\cite{2007PhRvD..76j4007A}, which can be viewed as a fourth-order finite difference 
scheme plus third-order adaptive dissipation. The dissipation coefficient is given by the 
maximum propagation speed in each grid point. For the fluid matter fields, we use a High 
Resolution Shock Capturing method with Monotonic-Centered limiter. The time evolution is 
performed through the Method of Lines using a third-order accurate Strong Stability 
Preserving Runge-Kutta integration scheme, with a Courant factor of $\Delta t/\Delta r = 
0.2$ so that the Courant-Friedrichs-Levy (CFL) condition dictated by the principal part of 
the equations is satisfied. Most of the simulations presented in this work have been done 
with a spatial resolution of $\Delta r = 0.00625 M_\odot$, in a domain with outer 
boundary situated at $r= 100 M_\odot$. We have verified, by 
changing the position of the outer boundary, that the results do not vary significantly 
with different choices of the boundary. We use maximally dissipative boundary 
conditions for the spacetime 
variables, and outflow boundaries for the fluid matter fields.

This nonlinear code has passed a large set of stringent tests. First, as it was shown 
in~Fig. 9 of Ref.~\cite{sus}, it already recovered the well-known frequencies of a 
neutron star with mass $M=1.4 M_{\odot}$ (as calculated, for instance,
in Ref.~\cite{2002PhRvD..65h4024F}). Furthermore, we have compared extensively our 
nonlinear and linear codes for $\bar {\cal C}=0$, finding an excellent agreement on the 
quasi-normal frequencies, as shown in Table~\ref{tau}.
\begin{table}
\begin{tabular}{c||cc}
 $M/R$ & $\tau_{\text{nonlinear}}$ & $\tau_{\text{linear}}$ 
\\ \hline\hline 
0.12 & 22.3  &  22.5 \\
0.14 & 22.4  &  22.3 \\
0.16 & 23.06 &  22.9 \\
0.18 & 24.94 &  25.2 \\
0.20 & 34.75 &  34.6
 \end{tabular}
\caption{ Characteristic oscillation times for $\bar {\cal C}=0$ and different 
compactness. We compare the result of the nonlinear simulations (second column) with 
those of the linear analysis (third column). 
}\label{tau}
\end{table}

\begin{figure}
\includegraphics[width=0.46\textwidth]{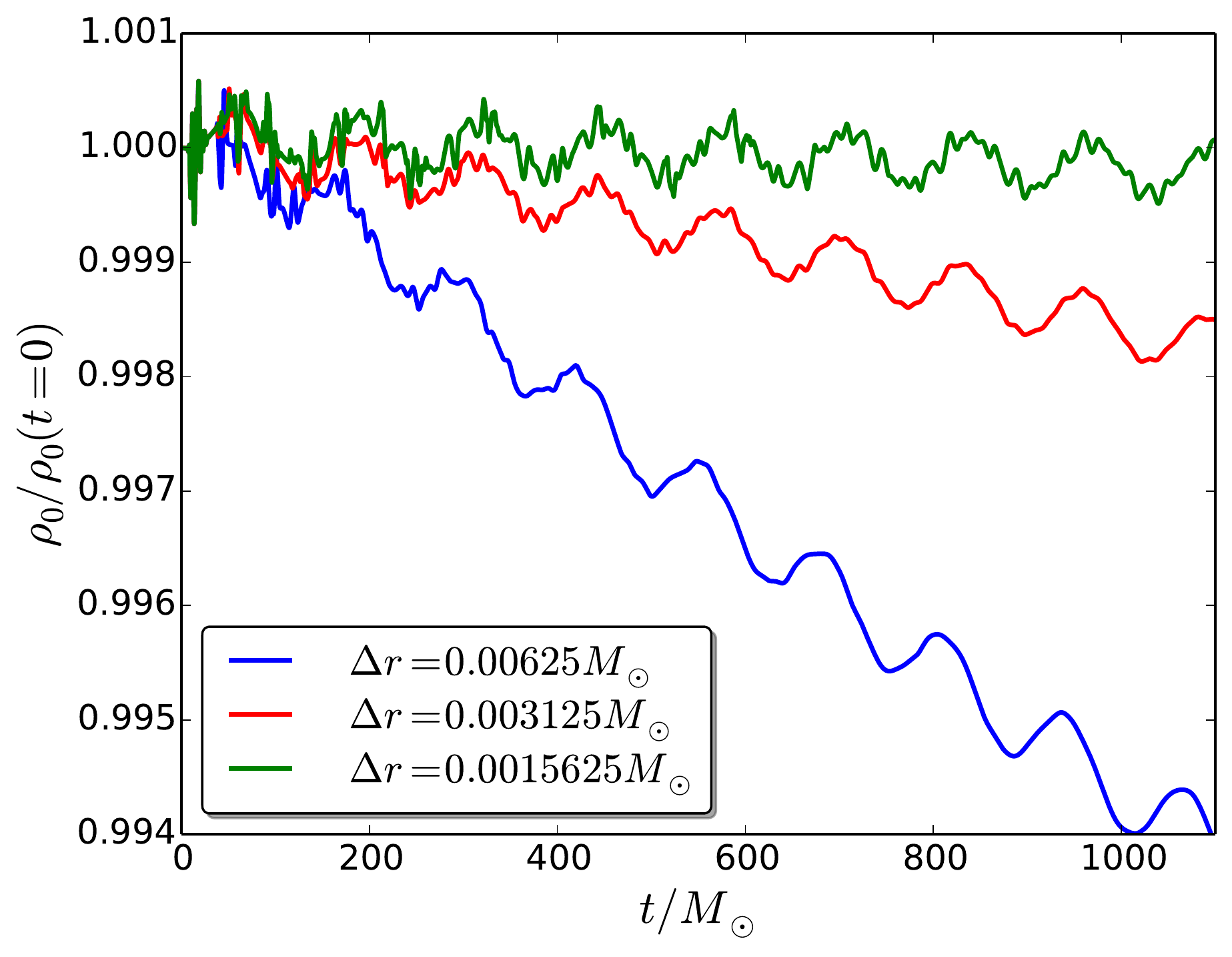} 
\includegraphics[width=0.46\textwidth]{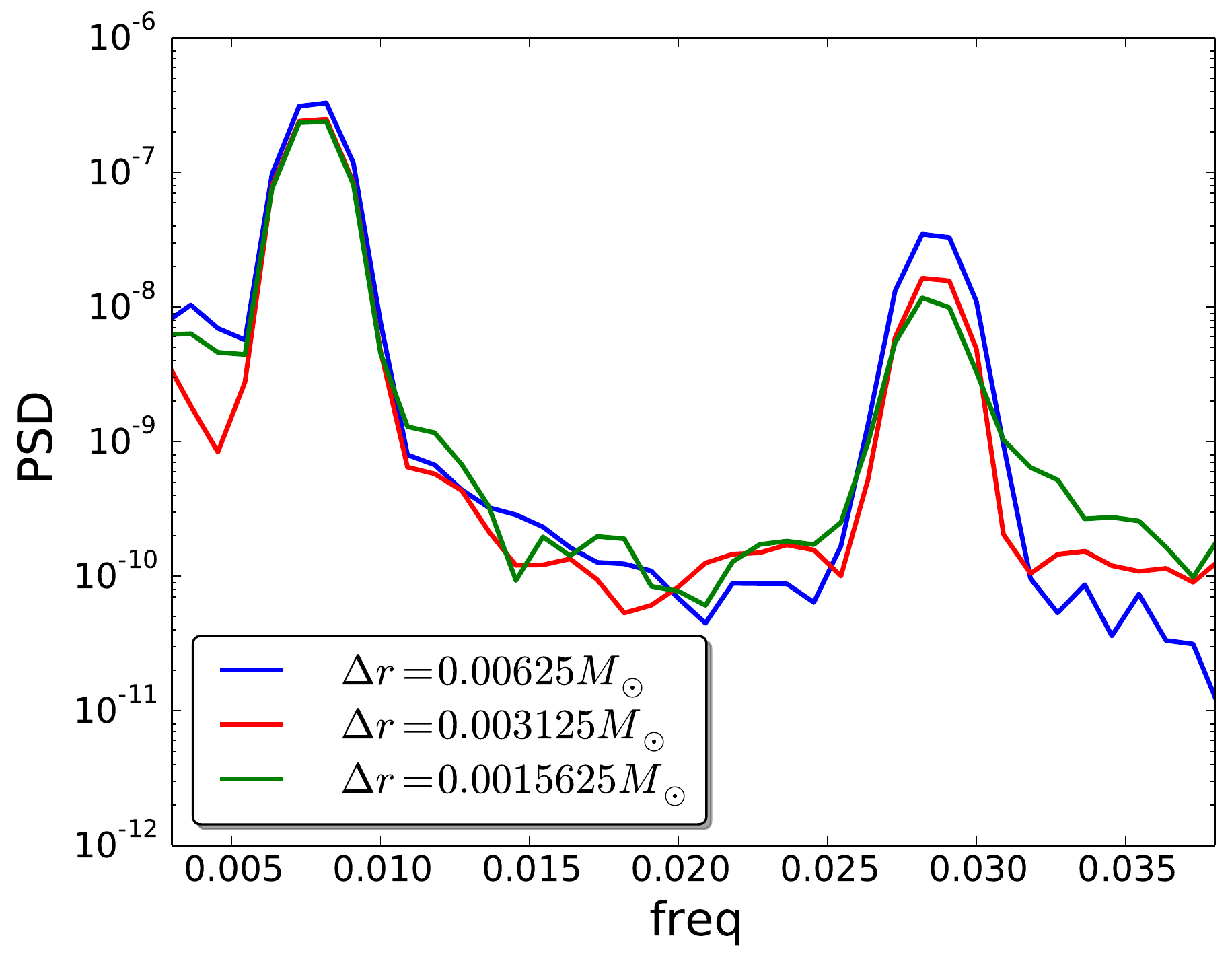} 
\caption{Convergence test for $\bar {\cal C}=1000$ and $M/R\approx0.306$. (Top) Central 
density $\rho_{0}$ as a function of time for three resolutions $\Delta r = \{0.00625, 
0.003125, 0.0015625\}M_\odot$.
(Bottom) Fourier transform of the central density.
The first peak is centered
at the same frequency (i.e., corresponding to the fundamental mode of the characteristic oscillation) for the three 
resolutions.
\label{rho0_convergence} }
\end{figure}

Finally, we always performed convergence tests, especially for the extreme cases with 
$\bar {\cal C}\gg1$. Indeed, in the latter cases passing convergence tests requires 
unusually high spatial/time resolutions to resolve steep structures near the radius of the 
star (see next section). For all stable cases we found that the results converge as 
expected. The frequencies reported in the main text are obtained with the highest 
resolution and we checked that they are almost insensitive to the time/spatial resolution.

We display the central density of the star for one of these convergence tests in 
Fig.~\ref{rho0_convergence}, corresponding to a configuration with $M/R=0.306$ with $\bar 
{\cal C} =1000$. 
We use $\Delta r = \{0.00625, 0.003125, 0.0015625\}M_\odot$ for this test.
The star is initially in equilibrium, only perturbed by numerical discretization errors, 
and oscillates with its associated normal frequencies. In addition, there is a 
deviation from the constant stationary value due to numerical errors, which decreases as 
the resolution is increased. The solution shows almost a second convergence. Notice that, 
although this value is below the third-order convergence expected for smooth solutions, is higher than the linear 
convergence expected in the presence of strong shocks.

\bibliographystyle{apsrev4}
\bibliography{Ref}

\end{document}